\documentclass[tikz,11pt,english,a4paper]{article}
\usepackage{float}
\usepackage{jcappub}
\usepackage[utf8]{inputenc}

\usepackage{bm}
\usepackage{algorithm}
\usepackage{nicefrac}
\usepackage{array}
\usepackage{multirow}

\usepackage[noend]{algpseudocode}
\makeatletter
\def\BState{\State\hskip-\ALG@thistlm}
\makeatother

\usepackage{fontawesome5}
\usepackage{hyperref}
\makeatletter
\newcommand{\github}[1]{%
   \href{#1}{\faGithub}%
}
\makeatother

\usepackage{lmodern}

\newcommand{\connect}{\textsc{connect}}
\newcommand{\class}{\textsc{class}}
\newcommand{\camb}{\textsc{camb}}
\newcommand{\montepython}{\textsc{MontePython}}

\newcommand{\polychord}{\textsc{PolyChord}}
\newcommand{\multinest}{\textsc{MultiNest}}
\newcommand{\aspic}{\textsc{aspic}}
\newcommand{\pyaspic}{\textsc{Pyaspic}}
\usepackage{siunitx}
\DeclareSIUnit \parsec {pc}

\usepackage[T1]{fontenc}

\usepackage{soul}

\usepackage{lmodern} 
\rmfamily 
\DeclareFontShape{T1}{lmr}{b}{sc}{<->ssub*cmr/bx/sc}{}
\DeclareFontShape{T1}{lmr}{bx}{sc}{<->ssub*cmr/bx/sc}{}

\begin{document}
\emergencystretch 3em


\title{Calculating Bayesian evidence for inflationary models using {\textsc{CONNECT}}}

\author[a]{Camilla T. G. Sørensen,}
\author[a]{Steen Hannestad,}
\author[a]{Andreas Nygaard,}
\author[a]{and Thomas Tram}

\affiliation[a]{Department of Physics and Astronomy, Aarhus University,
 DK-8000 Aarhus C, Denmark}

\emailAdd{camth@phys.au.dk}
\emailAdd{steen@phys.au.dk}
\emailAdd{andreas@phys.au.dk}
\emailAdd{thomas.tram@phys.au.dk}

\abstract{
Bayesian evidence is a standard tool used for comparing the ability of different models to fit available data and is used extensively in cosmology. However, since the evidence calculation involves performing an integral of the likelihood function over the entire space of model parameters this can be prohibitively expensive in terms of both CPU and time consumption. For example, in the simplest $\Lambda$CDM model and using CMB data from the Planck satellite, the dimensionality of the model space is over 30 (typically 6 cosmological parameters and 28 nuisance parameters). Even the simplest possible model requires $\mathcal{O}(10^6)$ calls to an Einstein--Boltzmann solver such as \class{} or \camb{} and takes several days.

Here we present calculations of Bayesian evidence using the \connect{} framework to calculate cosmological observables. We demonstrate that we can achieve results comparable to those obtained using Einstein--Boltzmann solvers, but at a minute fraction of the computational cost. As a test case, we then go on to compute Bayesian evidence ratios for a selection of slow-roll inflationary models.

In the setup presented here, the total computation time is completely dominated by the likelihood function calculation which now becomes the main bottleneck for increasing computation speed.
}

\maketitle

\section{Introduction}\label{sec:introduction}

Over the past three decades, a vast amount of cosmological data has yielded unprecedented knowledge of the physical model of our Universe. The standard $\Lambda$CDM model is described in terms of relatively few free parameters and provides a very good fit to almost all observational data. Various statistical techniques have been used to infer the value of the fundamental physical parameters of the model, including Bayesian parameter inference through marginalisation of the likelihood function (see e.g.~\cite{Lewis:2002ah,Torrado:2020dgo,Brinckmann:2018cvx}), and maximum likelihood techniques in the form of profile likelihoods (see e.g.~\cite{Holm:2023uwa,Karwal:2024qpt,Herold:2021ksg,Holm:2022kkd,Hannestad:2000wx}). Another extremely useful tool is the calculation of Bayesian evidence when comparing different models (see e.g.~\cite{bayesian} for a review). However, a major obstacle in evidence calculation is that it requires integration of the likelihood function over the entire prior volume, which, for high dimensional parameter spaces, can become prohibitively expensive.

Packages based on the nested sampling approach to likelihood integration~\cite{Skilling:2004pqw,Skilling:2006gxv} are by now available for carrying out such analyses in a relatively efficient manner. \polychord~\cite{Handley:2015fda} and \multinest{}~\cite{Feroz:2008xx} are among the most commonly used within the field of cosmology (see e.g.~\cite{Ashton:2022grj} for a recent review of methods and packages).
However, even with these packages, a reliable evidence calculation typically still requires millions of evaluations of the likelihood function. Each such evaluation requires running an Einstein--Boltzmann solver such as \class{}~\cite{Blas:2011rf} or \camb{} \cite{Lewis_2000} to calculate the relevant cosmological observables and takes on the order of tens of seconds on a single CPU core (although a significant speed-up can be achieved in cases where the model parameter space can be split in ``slow'' (cosmological) and ``fast'' (nuisance) parameters). This makes evidence calculations extremely expensive, both in terms of time and computational resources.

A way to mitigate this could be to use a cosmological emulator instead of the Einstein--Boltzmann solver code. Recent years has seen a surge in popularity of such emulators and they have been applied in many different ways. The most common kinds of emulators are either based on Artificial Neural Networks~\cite{SpurioMancini:2021ppk,Nygaard:2022wri,Gunther:2022pto,Bonici:2023xjk} or Gaussian Processes~\cite{Gammal:2022eob,Gunther:2023xhh}, both with their respective advantages and drawbacks. The applications range from standard Bayesian marginalisation to frequentist profile likelihoods~\cite{Nygaard:2023cus}, and Refs.~\cite{Piras:2024dml,Polanska:2024arc,mcewen2023machine} furthermore employed emulators to approximate Bayesian evidence using posterior samples and a modification of the harmonic mean estimator~\cite{NewtonRaftery}. While approximations of the Bayesian evidence are useful to roughly compare cosmological models with very different evidence, models that only differ slightly need better estimates (e.g. from nested sampling) in order to perform a meaningful comparison. Ref.~\cite{SpurioMancini:2021ppk} demonstrated that evidence computations could indeed be accurately computed using an emulator, albeit for the vanilla $\Lambda$CDM-model and using large-scale structure likelihoods and/or Planck Lite.

In this paper we test how the \connect{}~\cite{Nygaard:2022wri} framework fares on evidence calculations by performing Bayesian model comparison of a variety of different slow roll inflationary models using the publicly available \polychord{} package~\cite{Handley:2015fda}. Accurate profile likelihoods require the emulator to be very accurate around the region of best fit, but in general they do not require very accurate emulation of other regions in the parameter space~\cite{Nygaard:2023cus}. Marginalisation, on the other hand, requires integration over regions of parameter space. While this typically requires somewhat less precision around the absolute best fit, it requires the emulation to be reasonable over substantially larger regions. Evidence calculations are even more extreme in this regard since each evidence calculation requires integrating the likelihood function over the entire prior volume.

Given that evidence calculations are extremely time consuming due to the very large number of function evaluations required (typically millions of \class{} or \camb{} evaluations, each requiring tens of CPU core seconds), it is of substantial interest to investigate whether the \connect{} emulator can also be used for this purpose. In order to compare our results to model comparisons using standard Einstein--Boltzmann solvers, we use the same prior ranges and model parameterisations as in Ref.~\cite{planck2018}.

Finally, since we are using inflationary model selection as our test case, we must of course credit the pioneering work in Refs.~\cite{Ringeval:2013lea, Martin:2013nzq}. (See also Ref.~\cite{Martin:2024qnn} for a very recent update.) In these papers, the authors computed an effective likelihood by integrating out all non-inflationary parameters. A neural network was then trained to emulate this effective likelihood, allowing the authors to perform an exhaustive Bayesian model-comparison of most slow-roll inflationary models in the $\Lambda\text{CDM}$-model. Furthermore, we must also mention the early work done on inflationary model selection in Ref.~\cite{Easther:2011yq}.

The paper is structured as follows: In Section 2 we provide an overview of both the \connect{} framework and of Bayesian evidence calculations. Section 3 contains a description of how the \connect{} neural network emulator is constructed and validated using standard inflationary observables. Section 4 is then devoted to a description of how we implement the \aspic{} framework for describing slow-roll inflationary models and converting fundamental inflationary parameters to observables, and Section 5 contains our numerical results. Section 6 contains our runtime considerations. Finally, we provide our conclusions in Section 7.


\section{The {\bfseries\scshape connect} framework and Bayesian evidence}\label{sec:connect}

The \connect{} framework for emulation of cosmological observables has been tested extensively for cosmological parameter inference, using both Bayesian marginalisation and frequentist profile likelihoods~\cite{Nygaard:2022wri, Nygaard:2023cus}. 
A main advantage of \connect{} is that it contains both an emulator of cosmological observables as well as the framework needed to build and train the network. This means that emulators of new and non-standard cosmological models can easily be built and used to run cosmological parameter analyses within a single environment.
\connect{} trains a neural network based on training data sampled iteratively to best represent the likelihood function. This ensures that the neural network is most precise where the likelihood is large, which makes it ideal for parameter inference. The training data is gathered using the fast Planck Lite likelihood~\cite{2020}. The reason is that training requires a very large number of likelihood evaluations, which in the case of the full Planck likelihood would be prohibitively expensive. Because Planck Lite is somewhat less constraining than the full Planck likelihood, this gives us a set of training data that is more widely spread and this (along with a high sampling temperature, which guarantees adequate coverage of the training data set), yields a set of training data that can accurately represent several combinations of cosmological data sets (as long as either the full Planck likelihood, Planck Lite or similar CMB data is included) without the need to retrain the network. Furthermore, \connect{} is both the emulator as well as the procedure that trains the emulator. This makes it easy to quickly train a new physics model.

However, parameter inference as a statistical technique is designed for determining parameter values within a given model, assuming the model to be correct, i.e.\ it is not designed to qualitatively compare how well different models fare in fitting the available data. For this purpose other techniques, such as the Akaike information criterion in frequentist analysis (see e.g. Ref.~\cite{Liddle:2004nh}) or evidence in Bayesian analysis, are used instead. The Akaike information criterion relies on maximising the likelihood function and is therefore closely related to the profile likelihood technique already tested extensively with \connect{}~\cite{Nygaard:2023cus}. However, the Bayesian evidence calculation requires integrating the likelihood function over the entire prior volume, and testing the precision (and speed) with which \connect{} is able to perform this calculation is the main purpose of this work. 

The Bayesian evidence has been calculated with the code \polychord{}~\cite{Handley:2015fda,Handley:2015vkr}, which uses a version of nested sampling~\cite{Skilling:2004pqw} to calculate the evidence. The code is run from within the MCMC sampler \montepython{} \cite{Brinckmann:2018cvx, Audren:2012wb} with either \class{} or \connect{} as the cosmological theory code. We finally note that it is only necessary to train one model with \connect{}, because all the inflationary parameters for the different inflationary models can be mapped to the same physical parameters.

\section{Validation of \connect{} for evidence computation}\label{sec:validation}

A natural first step is to validate results for Bayesian evidence calculated using \connect{} versus brute force calculations based on \class{} (or \camb{}). 
The accuracy of \connect{} has been investigated thoroughly for both Bayesian parameter inference and profile likelihoods and found to be more than sufficiently accurate for such analyses, even in very extended parameter spaces (see~\cite{Nygaard:2022wri}). However, the calculation of Bayesian evidence typically lends more weight to regions in parameter space where the likelihood is only moderately good. This means that one cannot directly infer from these previous tests that \connect{} performs Bayesian evidence calculations at the required level of precision.

To test this, we calculate evidence in models based on $\Lambda$CDM, but with an extended inflationary component.  The basis is the simplest inflationary slow-roll approximation in which the primordial fluctuations are adiabatic, Gaussian, and purely scalar and can be parameterised using only the amplitude, $A_s$, and the spectral index, $n_s$. Beyond this, we have added the tensor-to-scalar ratio, $r$, as well as the effective curvature of the primordial spectrum, $\alpha_s$, so that the primordial fluctuation spectrum is fully described by four parameters: $A_s, n_s, r, \alpha_s$ 
\footnote{Validating \connect{} on this particular model has the advantage that since all the slow-roll models to be investigated can be mapped to the set of effective inflationary ``observables'', $A_s, n_s, \alpha_s, r$, we can infer that our set-up will also be valid for evidence computations using fundamental inflationary field parameters.}
in addition to the parameters needed to describe the content of the flat $\Lambda$CDM model: $\omega_b, \omega_\mathrm{cdm}, \theta_s, \tau_\mathrm{reio}$.

\begin{table}[t]
	\begin{center}
		\begin{tabular}{l c c}
			\hline\hline
			Parameter & Minimum value of prior & Maximum value of prior \\ [0.5ex]
			\hline
			$100 \times \omega_\mathrm{b}$ & $1.9$ & $2.5$ \\ [1ex]
			$\omega_\mathrm{cdm}$ & $0.095$ & $0.145$ \\ [1ex]
			$100 \times \theta_s$ & $1.03$ & $1.05$ \\ [1ex]
			$\ln10^{10}A_s$ & $2.5$ & $3.7$ \\ [1ex]
			$\tau_\mathrm{reio}$ & $0.01$ & $0.4$ \\ [1ex]
			$n_s$ & $0.94$ & $1.0$ \\ [1ex]
			$\alpha_s$ & $-0.3$ & $0.3$ \\ [1ex]
			$r$ & $0.0$ & $0.3$ \\ [1ex]
			\hline
		\end{tabular}
		\caption{The parameter bounds used to validate the results for Bayesian evidence calculated using \connect{} versus calculations based on \class{}. }
		\label{tab:valpar}
	\end{center}
\end{table}

The parameter bounds for $100 \times \omega_b$, $\omega_\mathrm{cdm}$, $100 \times \theta_s$, $\ln 10^{10} A_s$, and $\tau_\mathrm{reio}$ is the same as in Ref.~\cite{planck2018}. The bounds for these parameters as well as the bounds for $n_s$, $\alpha_s$, and $r$ can be seen in Table \ref{tab:valpar}.

Since our main goal is to demonstrate the feasibility of using the \connect{} framework for Bayesian evidence calculations, we will use the same data combination as in Ref.~\cite{planck2018}.
However, it should be stressed that evidence calculations are notoriously hard to compare because exact numbers are extremely sensitive to hyperparameter choices such as e.g.\ prior volume.
Therefore, it is not to be expected that a direct, quantitative comparison can be made between our results and those of \cite{planck2018}.
Our data sets therefore in all cases consist of the full Planck 2018 TT,TE,EE+lowE data~\cite{2020}, the Planck 2018 lensing data~\cite{planckVIII}, as well as the BICEP Keck 2015 data~\cite{Ade_2018}.

In the standard setting for \polychord{} when run through \montepython{}~\cite{Brinckmann:2018cvx} there is a distinction between ``slow'' (cosmological) and ``fast'' (nuisance) parameters. The \polychord{} wrapper for \montepython{} is hard-coded to use 0.75 of the total wall time of the computation for integration of the cosmological parameter space and 0.25 on the nuisance parameter space.
Given the difference in execution time between \class{} and the likelihood calls this typically leads to at least an order of magnitude more evaluation points in the nuisance parameter space than in the cosmological parameter space, but since the nuisance parameter space typically has much higher dimension the standard setting for \polychord{} with \montepython{} provides a reasonable division between the two sets of parameters.

\begin{table} [b]
	\begin{center}
		\begin{tabular}{p{0.2\linewidth} c c c}
			\hline \hline
			Case & $\log\mathcal{Z}$ & Likelihood calls & CPU hours used \\ [0.5ex]
			\hline
			\polychord{} with \class{} & $-1860.4 \pm 0.54$ & $1,419,152 / 115,988,382$ & 30,000 \\ [1ex]
			\polychord{} with \connect{} & $-1861.1 \pm 0.55$ & $1,569,491$ & 125 \\ [1ex]
			\hline
		\end{tabular}
		\caption{The Bayesian evidence calculated with \polychord{} using both \connect{} and \class{}. 300 live points were used, and the oversampling feature was turned off for \connect{}. Note that the log-evidence and the error estimates are comparable despite the \class{}-run using $\sim 100$ times more likelihood calls. The last column shows the CPU core-hours used in each \polychord{} run.}
		\label{tab:baye}
	\end{center}
\end{table}


However, when \polychord{} is run using \connect{} this division between parameter spaces becomes catastrophically wrong. The reason is that {\it all} function calls in this case takes the same time because CPU time is entirely dominated by the time taken for likelihood calls. This means that the nuisance parameter space becomes severely under sampled and only if a much larger number of live points is used can convergence be achieved. The solution to this problem is to let \polychord{} use its normal default setting in which all parameters are treated equally. In Appendix A we provide a more detailed discussion of the problem and its solution.

Using the new setting for \polychord{} with \connect{}, the Bayesian evidence for the above model is then calculated with \polychord{} using both \connect{} and \class{} using 300 live points in both cases.
The values of the evidences can be seen in Table \ref{tab:baye} together with the total number of likelihood calls in both cases. The resulting posteriors for the physical and inflationary parameters can be seen in Figure \ref{fig:val}.

In Appendix A we also discuss convergence in terms of the number of live points used. Although we do find that even using as little as 300 live points is enough to obtain robust results, the CPU requirements of the \connect{}-based runs are small enough that we opt to run all our inflationary model evidence calculations using 1200 live points.

\begin{figure} [tb]
	\centering
	\includegraphics[width=1.0\linewidth]{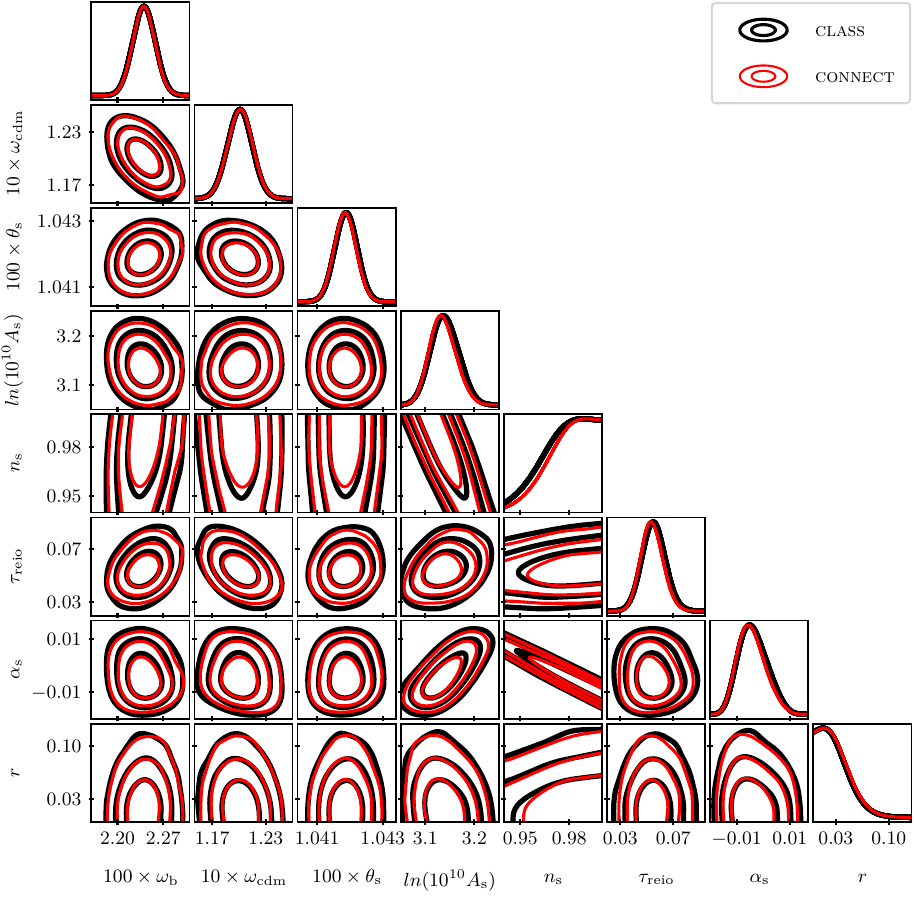}
	\caption{The posteriors for the physical and inflationary parameters with the bounds given in Table \ref{tab:valpar}. The contours correspond to 68.3\%, 95.5\%, and 99.7\% credible intervals.}
	\label{fig:val}
\end{figure}

\section{Inflationary model parameterisation}\label{sec:models}

In order to calculate Bayesian evidence for different inflationary models and their fundamental parameters, we have used the publicly available code \aspic{}~\cite{martin2023}. \aspic{} takes the inflationary model and its inflation parameters as input and calculates $n_s$, $\alpha_s$, and $r$, which can then be given as input to a neural network trained by \connect{}. The neural network then returns observables that can be used to compute a likelihood based on the given parameters of the inflationary model. Bayesian evidence is then computed using \polychord{}.

\aspic{} is written in \textsc{Fortran}, so in order to use the code with \connect{} and \montepython{}, we have written a Python wrapper, \textsc{PyAspic}\footnote{Available at \url{https://github.com/AarhusCosmology/PyAspic}.} for \aspic{} that can be called by \connect{}.

\begin{table} [t]
	\centering
	\addtolength{\leftskip} {-2cm}
	\addtolength{\rightskip}{-2cm}
	\begin{tabular}{l l l}
		\hline \hline
		\aspic{} model &  Model name in Ref.~\cite{planck2018} & Potential \\[0.5ex]
		\hline
		Higgs Inflation (HI) & $R+R^2/(6M^2)$ & \makebox[4cm][l]{\begin{tabular}[l]{@{}l@{}}\\[-2.3ex]$M^4 \left( 1-e^{-\sqrt{2/3}\phi/M_\mathrm{pl}} \right)^2$\\[0ex]\end{tabular}} \\[2ex]
		\hline
		Large Field Inflation (LFI$_2$) & Power-Law Potential & \makebox[1.75cm][l]{\begin{tabular}[l]{@{}l@{}}\\[-2.3ex]$M^4 \left( \frac{\phi}{M_\mathrm{pl}} \right)^{2}$\\[0ex]\end{tabular}} \\[2ex]
		\hline
		Large Field Inflation (LFI$_4$) & Power-Law Potential & \makebox[1.75cm][l]{\begin{tabular}[l]{@{}l@{}}\\[-2.3ex]$M^4 \left( \frac{\phi}{M_\mathrm{pl}} \right)^{4}$\\[0ex]\end{tabular}} \\[2ex]
		\hline
		Natural Inflation (NI) & Natural Inflation & \makebox[2.9cm][l]{\begin{tabular}[l]{@{}l@{}}\\[-2.3ex]$M^4 \left[ 1+\cos \left( \frac{\phi}{f} \right) \right]$\\[0ex]\end{tabular}} \\[2ex]
		\hline
		Loop Inflation (LI) & Spontaneously broken SUSY & \makebox[3.35cm][l]{\begin{tabular}[l]{@{}l@{}}\\[-2.3ex]$M^4 \left[ 1+\alpha \ln \left( \frac{\phi}{M_\mathrm{pl}} \right) \right]$\\[0ex]\end{tabular}} \\[2ex]
		\hline
		\makebox[5.3cm][l]{\begin{tabular}[l]{@{}l@{}}\\[-2.3ex]Colemann-Weinberg Inflation \\[-0.1ex] (CWI) \\[1ex]\end{tabular}} & Not in the reference & $M^4 \left[ 1+\alpha \left( \frac{\phi}{Q} \right)^4 \ln \left( \frac{\phi}{Q} \right) \right]$ \\[2ex]
		\hline
	\end{tabular}
	\caption{The inflationary models used in this paper. See the text for details on the parameters and their bounds}
	\label{tab:models}
\end{table}

The inflationary models used in this article have the following names in \aspic{}: Higgs Inflation (HI), Large Field Inflation (LFI) with $p=2$ and $p=4$, Natural Inflation (NI), Loop Inflation (LI), and Colemann-Weinberg Inflation (CWI). The models, their potentials, as well as their names in Ref.~\cite{planck2018} can be seen in Table \ref{tab:models}.

The bounds for the physical parameters ($100\times \omega_b$, $\omega_\mathrm{cdm}$, $100 \times \theta_s$, $\ln 10^{10} A_s$, and $\tau_\mathrm{reio}$) are the same as for the validation of the \connect{} network, and they can be seen in Table \ref{tab:valpar}. The inflation parameters and their bounds are $\ln \rho_\mathrm{reh}$ with bounds $\ln (1 \mathrm{ TeV})^4$ and $\ln \rho_\mathrm{end}$. The model NI has the parameter $f$ with bounds in logspace been given as $0.3 \leq \ln f \leq 2.5$, the model LI has the parameter $\alpha$ with bounds in logspace given as $-2.5 \leq \ln \alpha \leq 1.0$, and the model CWI has a parameter $\alpha$ held constant at $4e$ as well as the parameter $Q$ with bounds $0.00001 \leq Q \leq 0.001$. Furthermore, the model LFI also has a parameter $p$, and this model is run twice with p held constant at $p=2$ and $p=4$ respectively. The effective equation of state $w$ is $1/3$ for LFI with $p=4$ and $0$ for all other models.

\section{Numerical results}\label{sec:results}

\begin{table} [t]
	\begin{center}
		\begin{tabular}{l c}
			\hline \hline
			\aspic{} model & $\ln \mathcal{B}$  \\ [0.5ex]
			\hline
			Large Field Inflation (LFI$_2$) & $-8.8 \pm 0.4$  \\ [2ex]
			\hline
			Large Field Inflation (LFI$_4$) & $-51.2 \pm 0.4$ \\ [2ex]
			\hline
			Natural Inflation (NI) & $-4.6 \pm 0.4$  \\ [2ex]
			\hline
			Loop Inflation (LI) & $-4.7 \pm 0.4$ \\ [2ex]
			\hline
			Colemann-Weinberg Inflation (CWI) & $-19.7 \pm 0.6$  \\ [2ex]
			\hline
		\end{tabular}
		\caption{The calculated Bayesian evidence of the inflationary models with respect to the calculated Bayesian evidence for Higgs inflation. 
		In \cite{planck2018}, corresponding values for LFI$_2$, LFI$_4$, NI, and LI are -11.5, -56.0, -6.6, and -6.8 respectively. Coleman-Weinberg inflation was not tested in that work.
		The uncertainties on the values from Ref.~\cite{planck2018} is quoted as 0.3 in the article using 512 live points (note that estimated statistical uncertainties are typically significantly smaller for the same number of live points when using \class{} because of the very large number of likelihood evaluations in the nuisance parameter space).}
		\label{tab:results}
	\end{center}
\end{table}

After having validated the \connect{} framework for the purpose of calculating evidences, we now proceed to calculate evidence ratios for the selection of actual slow-roll models discussed in the previous section.
All the inflationary models given in Table \ref{tab:models} are run from \montepython{} with \polychord{}, \connect{}, and \aspic{}. The number of live points for the nested sampling algorithm is 1200 for all models \footnote{As discussed in Appendix A, even 300 live points is enough to calculate reliable evidences, but the calculation is sufficiently fast that we can use 1200 live points and thereby also achieve a somewhat smaller statistical uncertainty on the obtained results.}. The calculated evidence for all models with respect to the calculated Bayesian evidence for Higgs Inflation can be seen in Table \ref{tab:results}.

When comparing Bayesian evidence from different models, the \textit{Jeffreys scale} is often used~\cite{Jeffreys:1939xee}. Depending on the value of the Bayes factor between two models, the scale helps interpret if the strength of the evidence is either inconclusive, weak, moderate, or strong for one model compared to the other~\cite{bayesian}. The threshold values for Jeffreys scale can be seen in Table \ref{tab:jefscale}.

\begin{table}[b]
	\begin{center}
		\begin{tabular}{c c c l}
			\hline\hline
			$|\ln B|$ & Odds & Probability & Strength of evidence \\ [0.5ex]
			\hline
			<1.0 & <3:1 & <0.750 & Inconclusive evidence \\ [1ex]
			1.0 & $\sim$3:1 & 0.750 & Weak evidence \\ [1ex]
			2.5 & $\sim$12:1 & 0.923 & Moderate evidence \\ [1ex]
			5.0 & $\sim$150:1 & 0.993 & Strong evidence \\ [1ex]
			\hline
		\end{tabular}
		\caption{The strength of the Bayesian evidence interpreted by using the Jeffreys scale. The threshold values for the odds are 3:1, 12:1, and 150:1, which represents weak, moderate and strong evidence respectively. The table is taken from Ref.~\cite{bayesian}.}
		\label{tab:jefscale}
	\end{center}
\end{table}

Using Table \ref{tab:jefscale} to interpret the results given in Table \ref{tab:results}, it can be seen that Large Field Inflation with both $p=2$ and $p=4$ as well as Colemann-Weinberg Inflation are strongly disfavoured compared to Higgs Inflation. Natural Inflation and Loop Inflation both have a value of the Bayes factor that puts them right on the threshold between being moderately or strongly disfavoured compared to Higgs Inflation. Taking into consideration the uncertainty of $\pm 0.9$ for both models, it becomes impossible to put them into one category, and it is therefore concluded that the two models are moderately to strongly disfavoured compared to Higgs Inflation.

Comparing our results with Ref.~\cite{planck2018}, we find that they are in qualitative agreement regarding which models that are strongly disfavoured compared to Higgs Inflation. We note that the values for the Bayes factor found here are systematically more negative than the corresponding values in \cite{planck2018}, and in most cases deviate more than the estimated statistical uncertainty. We stress that this is {\it not} due to problems related to the \connect{} emulation, but is most likely related to small differences in the prior volume and/or parameterisation.

Ref. \cite{Martin:2024qnn} have also calculated the Bayesian evidence for different inflationary models using \aspic{} and a neural network, but they have trained their neural network on the effective likelihood, where all non-inflationary parameters have already been integrated out. They have used some different data sets than us, and the priors are not the same. But it is still possible to compare our results with theirs for two models: Large Field Inflation with $p = 2$ and Natural Inflation (even though their prior on $f$ is not identical to ours). They get $\ln \mathcal{B}_{\text{LFI}_2} = -7.35$ and $\ln \mathcal{B}_{\text{NI}} = -4.74$, which is in good agreement with our values seen in Table \ref{tab:results}.

\section{Runtime considerations}\label{sec:num}

The main reason for calculating Bayesian evidence with \connect{} instead of \class{} is that it is much faster even though we first have to train a neural network for the model. This can clearly be seen by comparing the time it took to calculate the Bayesian evidence in Section \ref{sec:validation} with \class{} and the time it took to train the neural network and calculate the evidence with \connect{} respectively. The calculation of the Bayesian evidence using \class{} took $\sim$30,000 CPU-hours on Intel Xeon E5-2680 v2 CPUs, whereas the calculation of the evidence with \connect{} (for $\Lambda$CDM+$\alpha_s$+$r$) took only $\sim$125 CPU-hours on Intel Xeon Gold 6230 CPUs. The difference in hardware might have a small effect, but it is most likely not more than a factor of $\sim$2. The training of the neural network (including sampling and calculation of training data) took $\sim$150 CPU-hours, so even with this included, the calculation of the Bayesian evidence is still much faster with \connect{} than with \class{}. Furthermore, the evidence for the different inflationary models all took less than $\sim$3500 CPU hours combined to calculate with \connect{} and 1200 live points, which is considerably less than what was required for $\Lambda$CDM+$\alpha_s$+$r$ with \class{} despite the inflationary models being more complicated as well as having 4 times as many live points.

When calculating the Bayesian evidence using \class{}, the dominant part of the calculation is the evaluation of \class{} itself. By using \connect{} instead, the evidence can be calculated without evaluating any of the hundreds of coupled differential equations in \class{}, and the limiting factor therefore becomes the Planck likelihood. To train the neural network using \connect{}, \class{} still needs to calculate the Einstein--Boltzmann equations, but the number of times \class{} is called during the training is much less than the number of times it is called when calculating the evidence without a neural network. When using \class{} to calculate training data for the neural networks, the total number of evaluations is $\sim$50,000, which is 30 times fewer evaluations than the \polychord{} run using \class{}. 

In summary, performing an evidence calculation using \class{} takes $X\sim 30,000 \text{CPU-hours}$, while an evidence calculation using \connect{} requires two steps: First, training data is generated iteratively and the neural network is trained and then the evidence calculation proceeds. The first step takes $X_1 \sim  150\text{CPU-hours}$, while the second step takes $X_2 \sim  125\text{CPU-hours}$. Because the model can be reused for different inflationary models and different dataset combinations, the combined runtime for $m$ models and $d$ dataset combinations will be $X_\text{tot} = m\times d \times X$ when using \class{} but only $X_\text{tot} = X_1 + m\times d \times X_2$ when using \connect{}. Thus, if we consider the case were we are computing Bayes factors for a handful of models with a few dataset combinations, the training time becomes completely negligible, and \connect{} delivers a speedup factor of $\sim240$ compared to \class{}.

\section{Discussion and conclusions}\label{sec:conclusion}

We have tested the use of the \connect{} framework for calculating Bayesian evidences in cosmology using inflationary models as a test case. \connect{} has previously been shown to emulate cosmological observables at a level of precision more than adequate for performing Bayesian parameter inference and for computing profile likelihoods. However, since the calculation of Bayesian evidence typically puts more weight on regions of parameter space in which the likelihood is only moderately good it cannot {\it a priori} be assumed that \connect{} delivers suitable precision for this task.

Using the standard set of ``observational'' parameters describing slow-roll inflation models, $A_s, n_s, \alpha_s, r$, we found that running \polychord{} with default settings through \montepython{} leads to severe undersampling of the nuisance parameter space when we use \connect{} rather than \class{}. We traced this problem to a default setting in the \polychord{} wrapper which splits parameters into ``slow'' (cosmological) and ``fast'' (nuisance) parameters, and devotes 0.75 of the wall time to sampling the slow parameter space. When running \polychord{} with \class{} this leads to a suitable division of labour between slow and fast parameters. However, when run with \connect{} it leads to the mentioned under sampling of nuisance parameters and poor convergence of the computation. In fact, the \connect{}-based runs typically required an order of magnitude more live points to achieve the same precision as the \class{}-based runs.

To fix the problem we ran \polychord{} with all parameters treated equally (i.e.\ no splitting into ``slow'' and ``fast'' parameters) and found that results become compatible with \class{}-based results with the same number of live points, thus validating that \connect{} can replace the use of \class{} for evidence computations. This in turn reduced runtime tremendously with the evidence calculations now being completely dominated by the likelihood calls.


Having validated the \connect{} framework for this purpose we then proceeded to calculate Bayesian evidence for a number of slow-roll inflationary models by using the \aspic{} library to convert inflationary parameters to observable inflationary parameters. We found evidence ratios between models very similar to those reported in Ref.~\cite{planck2018} and in all cases within the same evidence strength brackets. Furthermore, all the calculations of the Bayesian evidence was done with 1200 live points and on 24 tasks with 1 CPU for each task, and the calculations were done within 24 hours. Using a neural network therefore drastically reduces the runtime for these calculations, making it possible to easily use Bayesian evidence as a tool to compare different theoretical models.

Based on the tests carried out and presented here we are therefore confident that \connect{} can be used for calculations of Bayesian evidence in cosmology, vastly reducing the often prohibitive runtimes of such calculations. This will make it possible to start using Bayesian evidence as a tool in theoretical cosmology. Right now, theoretical cosmology is mostly done using Bayesian and frequentist parameter estimation, but it will now be possible to also use Bayesian evidence and thereby compare how good one cosmological model is compared to another cosmological model.

\vspace*{1.5cm}

\noindent {\bf Reproducibility.}
We have used the publicly available \connect{} framework available at \url{https://github.com/AarhusCosmology/connect_public/tree/2406.03968} to create training data and train neural networks. To calculate the Bayesian evidence, we have used the publicly available program \polychord{} available at \url{https://github.com/PolyChord/PolyChordLite} as well as the program \montepython{} publicly available at \url{https://github.com/AarhusCosmology/montepython_public/tree/2406.03968}. Lastly, we have used the program \aspic{} to calculate the inflationary models and their fundamental parameters. This has been done with the publicly available \textsc{Python} wrapper \pyaspic{} available at \url{https://github.com/AarhusCosmology/PyAspic}. \aspic{} is publicly available at \url{http://cp3.irmp.ucl.ac.be/~ringeval/upload/patches/aspic/}.

\section*{Acknowledgements}
We thank J\'{e}r\^{o}me Martin, Christophe Ringeval, and Vincent Vennin for valuable comments on the manuscript, and Will Handley for fruitful discussions on \polychord{}. Furthermore, we acknowledge the use of computing resources from the Centre for Scientific Computing Aarhus (CSCAA). AN and TT was supported by a research grant (29337) from VILLUM FONDEN. CS and SH were supported by a grant from the Danish Research Council (FNU).

\section*{Appendix A}

\begin{figure} [tb]
	\centering
	\includegraphics[width=1.0\linewidth]{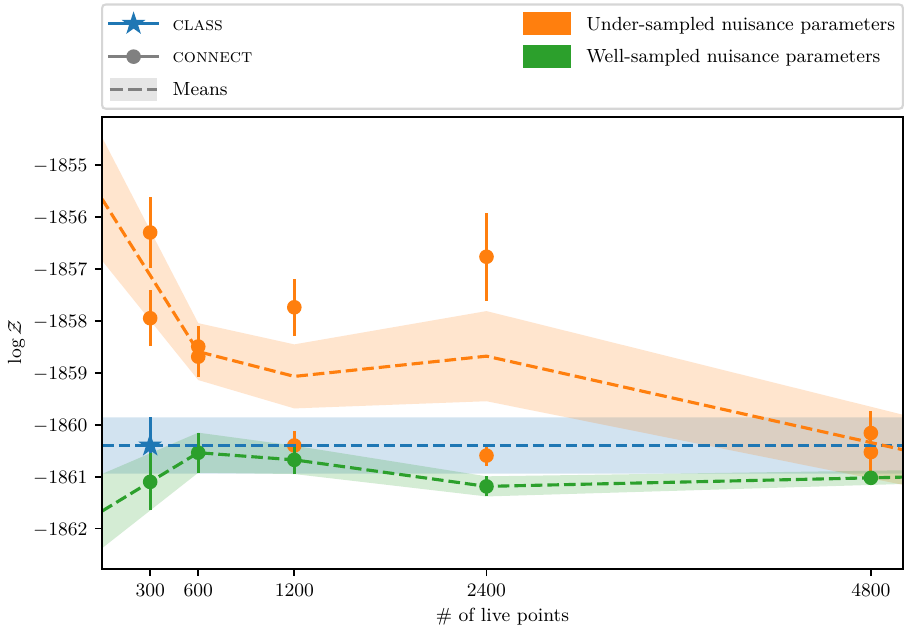}
	\caption{Evidence calculation of the the phenomenological $(A_s, n_s, \alpha_s, r)$-model using \polychord{}. We compare \class{}, \connect{} with default \montepython{}, and \connect{} with our corrected \montepython{}. Without the fix, 4800 live points are needed to obtain a converged result, whereas \class{} already seems converged when using 300 live points. With the fix to \montepython{}, \connect{} is in agreement with \class{} and is no longer biased.}
	\label{fig:evidencebiasfix}
\end{figure}

In this appendix we validate the use of \polychord{} with \connect{} and discuss convergence in terms of the number of live points. As we discussed in Section \ref{sec:validation}, the default setting for \polychord{} in \montepython{} leads to severe undersampling of the nuisance parameter space when using \connect{} 
\footnote{There are other situations where the default behaviour can be sub-optimal, see e.g. the issue at \url{https://github.com/brinckmann/montepython_public/issues/374}.}.
This undersampling leads to a bias in the ensemble mean log-evidence, unless a very large number of live points is being used, as shown in Figure~\ref{fig:evidencebiasfix}.


\begin{table} [tb]
	\begin{center}
		\begin{tabular}{l c c}
			\hline \hline
			Case & Bayesian evidence ($\log\mathcal{Z}$) & ``slow''/``fast'' likelihood calls  \\ [0.5ex]
			\hline
			\polychord{} with \class{} & $-1907.4 \pm 0.92$  &  398,478 / 25,689,051 \\ [1ex]
			\polychord{} with \connect{} & $-1898.2 \pm 1.43$ & 469,336 / 329,299 \\ [1ex]
			\hline
		\end{tabular}
		\caption{The Bayesian evidence, as well as the number of likelihood evaluations, for the LFI$_4$ model calculated with \polychord{} using both \connect{} and \class{}, both using the standard \montepython{} settings for \polychord{} in which parameters are split in ``slow'' and ``fast'' categories and 0.75 of the total wall time is spent integrating the slow parameter space. This test run was performed using 100 live points.
}
		\label{tab:ncalls}
	\end{center}
\end{table}

That the nuisance parameter space becomes under sampled with standard settings is very evident from Table \ref{tab:ncalls} in which it can be seen that even though the number of evaluations in the slow parameters are comparable in the two cases, the number of evaluations in the fast parameters are a factor of 80 smaller when using \connect{}.

Once diagnosed this problem can be easily fixed by disabling the oversampling feature in \texttt{PolyChord.py} and treating all variables democratically. 
Running \polychord{} with \connect{} using these settings produces a Bayesian evidence of $\log\mathcal{Z} = -1906.2 \pm 0.9064$ using a total of 2,035,411 likelihood evaluations.

In order to further test convergence of \polychord{} with both standard and ``new'' settings we have performed a series of test runs for the 
the phenomenological $(A_s, n_s, \alpha_s, r)$-model, varying the number of live points. The results are shown in Figure~\ref{fig:evidencebiasfix} from which we can conclude that \connect{} with standard \polychord{} settings requires (at least) 4800 live points to achieve the same precision as \class{}-based runs with 300 live points. With the fix in place the \connect{}-based runs converge as quickly as the \class{}-based runs in terms of number of live points, but using a smaller total number of likelihood evaluations. 

\bibliographystyle{utcaps}
\bibliography{evidence2023}

\end{document}